
%
%
\magnification=\magstep1

\def\picture #1 by #2 (#3){
\vtop{\vskip #2
\special{picture #3}
\hrule width #1 height 0pt depth 0pt
\vfil}}
\def\scaledpicture #1 by #2 (#3 scaled #4){
\dimen0=#1 \dimen1=#2
\divide\dimen0 by 1000 \multiply\dimen0 by #4
\divide\dimen1 by 1000 \multiply\dimen1 by #4
\picture \dimen0 by \dimen1 (#3 scaled #4)}
\hoffset=.4truecm
\voffset=.5truecm
\hsize=15truecm
\vsize=23truecm
\baselineskip=12pt
\hfuzz=18pt
{\nopagenumbers
\vbox{\vskip 2truecm
\centerline{\bf TOPOLOGICAL MODELS ON THE LATTICE AND}
\vskip 0.1cm
\centerline{\bf A REMARK ON STRING THEORY CLONING}}
\vfil
\vfil
\centerline{C. BACHAS}
\vskip 0.2cm
\centerline{\it Centre de Physique Th\'eorique, Ecole
Polytechnique\footnote{$^{\dag}$}{\sevenrm Laboratoire
Propre du Centre National de la Recherche Scientifique UPR
A.0014}}
\centerline{\it 91128 Palaiseau Cedex, France}
\vskip 0.3cm
\centerline{and}
\vskip 0.3cm
\centerline{P.M.S. PETROPOULOS}
\vskip 0.2cm
\centerline{\it Service de Physique Th{\'e}orique de
Saclay\footnote{$^{\ddag}$}{\sevenrm Laboratoire de la
Direction des Sciences de la Mati{\`e}re du
Commissariat {\`a} l'Energie Atomique}}
\centerline{\it
91191 Gif-sur-Yvette Cedex, France}
\vfil
\centerline{\bf Abstract}
\vskip 0.6cm
The addition of a topological model to the matter content
of a conventional closed-string theory leads to the
appearance of many perturbatively-decoupled space-time
worlds. We illustrate this by classifying topological
vertex models on a triangulated surface. We comment on
how such worlds could have been coupled in the Planck
era.
\vfil
\noindent
P.A.C.S.:  05.50.+q, 11.15.-q, 11.17.+y
\par\noindent CPTH-A158.0392\par\noindent
\line{SPhT/92-038  \hfil March 1992}
\eject}
\pageno=2
\vskip 0.6cm
\noindent{\bf 1. Many worlds in string theory}
\vskip 0.3cm
Topological quantum field
theories [1, 2] are characterized by
their invariance under local smooth deformations of the
background metric.
Thus adding a  two-dimensional
topological model to the matter content
of a conventional critical closed-string theory should not
affect the decoupling of the Liouville mode and hence also
the theory's consistency. Could this then imply that
critical string theory is not unique?
\vskip 0.3cm
In order to address this question we
must specify more precisely what we mean by
topological models. One way to define them, following
Atiyah [3], is through a
set of axioms. The basic data is a
finite-dimensional space  ${\cal H}$
of states
created by local field operators
$\{\phi _{1}\equiv {\bf I}, \phi _{2}, \ldots ,\phi
_{M}\}$, together with their (symmetric)
two- and three-point functions on the sphere:
$$
\eqalign{\langle \phi _{a} \phi _{b} \rangle &
_{\rm sph}= \, \eta _{ab} \cr
\langle \phi _{a} \phi _{b} \phi _{e}&\rangle
_{\rm sph}= \, c_{abe} \; .\cr
}
                                       \eqno (1)
$$
The two-point function
$\eta _{ab}\equiv c_{1ab}$
must define
a non-singular bilinear
inner product, which identifies ${\cal H}$
with its dual:
$(\phi _{a})^{\star } \equiv  \phi ^{a}=\eta ^{ab}\, \phi
_{b}$.
Here and in the
sequel indices are raised with the inverse metric
$\eta ^{ab}$ and
repeated indices are implicitly summed. Unitarity
requires the correlation functions of
self-adjoint operators to be real. Using the three-point
functions and the metric,  we can give
${\cal H}$
the structure of a {\it commutative} operator algebra
$$
\phi_{a}^{\phantom e} \times \phi_{b}^{\phantom e}
= c_{ab}^{\; \; \; e} \,\phi_{e}^{\phantom a}
                                     \; .  \eqno (2)
$$
Now the crucial axiom is the assumption
of {\it factorization}, which allows us to calculate
a  correlation function on an arbitrary Riemann surface by
first deforming the surface into a  collection of
three-punctured spheres connected by long thin tubes,
then cutting the latter by inserting the identity
operator  ${\bf I} =
\vert \phi _{a}\rangle \langle \phi^{a}\vert
$.
Factorizing, in particular, the four-point
function on the sphere along two different channels
gives the duality constraint
$$
c_{ab}^{ \; \; \; e}\,
c_{ec}^{ \; \; \; d}  =
c_{bc}^{ \; \; \; e} \,
c_{ae}^{ \; \; \; d}
                                    \;,     \eqno(3)
$$
which implies that the commutative operator algebra is
also {\it associative}. Under
these conditions it can be shown that there exists a
self-adjoint basis,
{${\tilde \phi}_{a}^{\phantom {\}}} =
(S^{-1})_{a}^{\; \, b}\, \phi_{b}^{\phantom {\}}}$},
in
which the algebra is diagonalized:
$$
\tilde \eta _{ab}=\delta _{ab}\; \; {\rm and}\; \;
{\tilde c}_{abe} = \cases{\lambda _{a} &if $a=b=e \; ,$\cr
0  &otherwise.\cr}
                                                \eqno(4)
$$
The non-vanishing correlation functions in this basis
are
$$
\left\langle \tilde \phi _{a}^{n}
\right\rangle _{\Gamma }^{\phantom I}=
\lambda _{a}^{2\Gamma -2+n}
                             \; ,            \eqno (5)
$$
where $\Gamma$ is the genus of the Riemann surface. When
translated in the original basis these read
$$
\left\langle \phi _{a_{1}}^{\phantom a} \ldots
\phi _{a_{n}}^{\phantom a}
\right\rangle _{\Gamma }^{\phantom I}=\sum_{a=1}^{M}
S_{\; \, a_{1}}^{a}\ldots S_{\; \, a_{n}}^{a} \,
\, \lambda _{a}^{2\Gamma -2+n}
                                \; .     \eqno (6)
$$
Note that the requirement
$\langle {\bf I} \, \phi_a \phi_b\rangle_{\rm sph} =
\eta_{ab}$
implies that
$\lambda _{a}^{\phantom 1}=\left(S_{\; \,1}^{
a}\right)^{-1}$.
\vskip 0.3cm
Suppose now that we tensor such a topological model with
the matter content of a conventional critical
closed-string theory. Both the
$SL(2,{\bf C})$
vacuum and all physical vertex
operators will in this case carry an extra index
$a=1,\ldots ,M$,
since multiplication by $\phi _{a}$ does not change the
conformal properties of fields.
The string amplitudes are simply
those of the conventional parent  string theory,
multiplied by the
(constant in moduli space) correlation functions of the
topological model\footnote{$^{(*)}$}{\sevenrm What we are
here discussing is, $\scriptstyle a$ $\scriptstyle priori$,
simpler
than the so called "topological strings"  [2,~4], obtained
by coupling topological matter to
$\scriptstyle  topological$ $\scriptstyle
gravity$. Indeed, the contact algebra of topological
strings does not seem to factorize into the contact
algebra of pure gravity times the operator algebra of
topological matter. To be sure, the structure of these
models is  not yet fully elucidated.}. We thus obtain $M$
copies of the graviton, dilaton, antisymmetric tensor, {\it
etc} \dots, which at first sight may appear to interact.
This is, however, an illusion since in the "tilde"
basis  these copies decouple to all orders in the
string loop expansion. A similar phenomenon has been
observed before, in the continuum
limit of matrix models [5]. Note that these $M$ copies
differ here only in the value of the string coupling
constant {\it i.e.} the vacuum expectation value of the
dilaton field, but other backgrounds can be also varied
independently. For instance, modding out by a reflection
of some internal coordinate times a ${\bf Z }_2$ symmetry
of the topological model will yield a collection of string
theories defined either on the circle or the orbifold.
\vskip 0.3cm
What we see here is another entry in the long dictionnary between world-sheet
and
space-time properties. A {\it topological
model} on the former translates into the
appearance of {\it many worlds} in the latter. If
these
worlds were truly decoupled, this remark would have only
philosophical value. However, in a theory of gravity,
such
worlds could have been  coupled in the Planck era
and/or through non-perturbative
effects, as illustrated in fig.~1. Understanding such
effects in string theory is therefore intimately
connected with understanding how a topological phase of
two-dimensional matter can be reached. Furthermore,
as has been argued by Witten,  such understanding
may be also relevant in the study of space-time
singularities~[6], as well as of a possible phase of
unbroken general  coordinate invariance [1, 7, 13].
\vskip 0.3cm
To make further progress on these issues we must, however,
abandon the above
axiomatic definition of topological models and study
instead how these arise from
 some local world-sheet dynamics. Several possibilities
have been suggested in the
literature: $\sigma $-models on a manifold with complex
structure [1, 2, 8], twisted $N=2$
supersymmetric models  [2, 9], gauged WZW models
[10] and finally topological
models on a cut-off triangulated surface
[11,~12]. The absence of spurious degrees of freedom for
the metric makes the characterization of
lattice topological models particularly simple and these
models will be the subject of the remainder of this
letter. We will, in particular, show how under some
assumptions they can be completely classified, confirm the
above cloning of string theories and suggest a
qualitative but intuitively appealing picture for the
wormhole of fig.~1. Though many parts of our analysis have
appeared in the literature in various contexts before,
putting them in a new perspective could, we hope, be a
useful prelude towards addressing the aforementionned hard
non-perturbative issues of string theory.
\vskip 0.6cm
\noindent {\bf 2. Topological models on a triangular
lattice}
\vskip 0.3cm
Consider an oriented genus-$\Gamma $ surface
${\cal M}_{\Gamma }$
made out of
$A$
identical equilateral triangles. As is well known, the way
of gluing these triangles together encodes all invariant
information about the underlying two-dimensional metric
[14]. We will consider a class of ("matter") spin models
on  ${\cal M}_{\Gamma }$ defined as follows: the
spins, denoted by  lower-case Greek letters
$\{\alpha, \beta ,\ldots =1,\ldots,s\}$,
live on
the oriented links of the lattice.
To every oriented triangular plaquette with spins
$\alpha, \beta $ and $\gamma$
we assign a Boltzmann weight
$P_{\alpha\beta\gamma}$,
while to every link with spins
$\alpha$ and $\beta$
in the two orientations we assign a weight
$\ell^{\alpha\beta}$. Both
$P_{\alpha \beta \gamma }$ and $\ell ^{\alpha \beta }$
must be symmetric under cyclic permutations of their
indices, which corresponds to local rotations on the
surface. Plaquette  weights, on the other hand,
need not be invariant under  orientation change, so that
in general
$P_{\alpha\beta\gamma}$ is not equal to
$P_{\alpha\gamma\beta}$.
\vskip 0.3cm
The partition function is a product
of plaquette and link weights, summed over all possible
values of the spins on the oriented links of the lattice.
More general correlation functions can be defined by
drilling holes on the surface and fixing the values of the
spins on their boundaries. As a simple illustration
consider two triangles glued together to form a disk or a
cylinder as shown in fig. 2. The corresponding
correlation functions, that would be equal if the
plaquette weights were fully symmetric, read:
$$
D_{\alpha \beta }^{\phantom \gamma }=
P_{\alpha \rho \sigma }^{\phantom \gamma }\,
P_{\beta \sigma '\rho '}^{\phantom \gamma }\,
\ell _{\phantom \beta }^{\sigma \sigma '}\,
\ell _{\phantom \beta }^{\rho \rho '}
\equiv \,
P_{\alpha \rho }^{\; \; \;  \; \sigma }\,
P_{\beta \sigma }^{\; \;  \; \; \rho }
                                       \eqno (7)
$$
and
$$
C_{\alpha \beta }^{\phantom \gamma }=
P_{\alpha \rho \sigma }^{\phantom \gamma }\,
P_{\beta \rho '\sigma '}^{\phantom \gamma }\,
\ell _{\phantom \beta }^{\sigma \sigma '}\,
\ell _{\phantom \beta }^{\rho \rho '}
\equiv \,
P_{\alpha \rho }^{\; \; \;  \; \sigma }\,
P_{\sigma \beta }^{\; \;  \; \; \rho }
                              \; ,         \eqno (8)
$$
where indices are here raised with
$\ell^{\alpha\beta}$.
All  partition
(correlation) functions are
clearly invariant (covariant) under the similarity
transformation
$$
\eqalign{
\ell^{\alpha \beta }_{\phantom \gamma }\to \;
T_{\; \, \gamma }^{\alpha }\,
T_{\; \, \delta }^{\beta }\,
&\ell^{\gamma \delta }_{\phantom \gamma }\cr
P_{\alpha \beta \gamma \phantom {\}}}^{\phantom {)}}\to
\;
\left(T^{-1}\right)^{\; \; \delta }_{\alpha }\,
\left(T^{-1}\right)^{\; \; \epsilon }_{\beta }\,
&\left(T^{-1}\right)^{\; \; \zeta }_{\gamma }\,
P_{\delta \epsilon \zeta \phantom {\}}}^{\phantom {)}}
\; ,\cr }
                                       \eqno (9)
$$
with $T$ an arbitrary invertible complex matrix. These
transformations
 define an equivalence relation
among different spin models. A further equivalence can
be defined through {\it restriction}, if in some basis
certain values of the spin never occur in the interior of
${\cal M}_{\Gamma }$. If, for example,
$\ell^{\alpha\beta}$ were degenerate we could obtain an
equivalent theory by restricting the values of the spins
to those labelling a basis for the subspace
of non-null eigenvectors. Without loss
of generality we may therefore assume in what follows that
$\ell^{\alpha\beta}$ has an inverse,
which we denote by $\ell_{\alpha\beta}$. Note finally
that the models considered here are the most general
vertex models on the dual
$\varphi^3$-graph ${\cal M}^*_{\Gamma}$. We will
refer to them for brevity as {\it vertex models}.
\vskip 0.3cm
Let us consider now the behaviour of correlation functions under local
variations of
the metric in the interior of ${\cal
M}_{\Gamma }$. The key observation
[11,~12,~15] is that these latter can be generated
by two elementary moves: the link-flip and pyramid moves
illustrated in
figs 3({\it a}) and ({\it b}) respectively.
We may therefore {\it define the class of topological
vertex models by  imposing invariance under
these two moves.} The corresponding conditions read:
$$
P_{\alpha \beta}^{ \; \; \; \; \epsilon}\,
P_{\epsilon \gamma}^{ \; \; \; \; \delta}  =
P_{\beta \gamma}^{ \; \; \; \; \epsilon} \,
P_{\alpha\epsilon}^{ \; \; \; \; \delta}
                               \eqno(10{\rm a})
$$
and
$$
P_{\alpha \beta \gamma }^{\phantom \gamma }=\,
P_{\alpha \delta }^{\; \; \; \; \zeta }\,
P_{\beta \epsilon }^{\; \; \; \; \delta }\,
P_{\gamma \zeta }^{\; \; \; \; \epsilon }
                          \; .   \eqno (10{\rm b})
$$
When these are satisfied,
correlation functions only depend on the genus of the
surface and on the values of the spins at its boundaries.
We will therefore denote them by
$C_{\left\{\alpha ^{(1)}\right\}\cdots \left\{\alpha
^{(n)}\right\}}^{\, (\Gamma )}$
where
$\left\{\alpha ^{(h)}\right\}$
are the values of the spins ordered according to the
induced  orientation around
the $h\, $th hole. A special name will be reserved for
correlation functions on the sphere with
$n$ length-one boundaries, or one length-$n$ boundary.
We will refer to them for short as $n$-point functions on
the {\it sphere} and {\it disk}, and use
the already anticipated notation:
$C_{\alpha _{1}\ldots \alpha _{n}\phantom{\}}}^{\phantom)}
\! \! \! \equiv
C_{\{\alpha _{1}\}\cdots  \{\alpha _{n}\}}^{\, (\Gamma
=0)}$
and
$D _{\alpha _{1}\ldots
\alpha _{n}\phantom{\}}}^{\phantom)} \! \! \! \equiv
C_{\{\alpha _{1}
\ldots \alpha _{n}\}}^{\, (\Gamma =0)} $
respectively. Note that the three-point
function on the disk is simply the plaquette weight:
$P_{\alpha \beta \gamma}\equiv D_{\alpha \beta \gamma}$.
\vskip 0.3cm
Some comments are in order here concerning the above
definition of topological models. First, it
includes as special cases all models studied in refs
[11,~12]. Nevertheless, it could be still conceivably
relaxed in a variety of ways.  One may, for instance,
demand topological invariance only in the continuum
($A\to \infty$) limit, or for only a subset of
external (boundary) states. One may also drop the
pyramid-move condition altogether. This introduces only
area in addition to genus dependence,  because link flips
suffice by themselves to connect any two surfaces of fixed
$A$ and $\Gamma $ to each other [15]. Finally one may
consider continuous and/or unbounded spins. We will
comment on some of these variations below, though an
exhaustive study lies beyond the scope of the present
letter.
\vskip 0.3cm
The conditions for
topological invariance, eqs (10a, b), have a simple
interpretation, if we define a formal algebra ${\cal A}$
generated by a basis of linearly independent vectors
$\left\{\varpi _{\alpha } \; , \; \; \alpha =1,\ldots,
s\right\}$,  with
multiplication rules
$$
\varpi_{\alpha}^{\phantom \gamma } \times
\varpi_{\beta}^{\phantom \gamma } =
P_{\alpha\beta}^{ \; \; \; \; \gamma} \,
\varpi_{\gamma}^{\phantom \gamma } \; .\eqno (11)
$$
The link-flip condition  implies that the
algebra is {\it associative}. In that case, the structure
constants define the so called {\it regular
representation} where
$\varpi_{\alpha }^{\phantom \gamma }$ is represented
by a matrix with entries
$P_{\alpha\beta}^{ \; \; \; \; \gamma}$,
$\beta $ and $\gamma $ being the column and row indices
respectively. Both eq. (7) and the pyramid-move condition
(10b) are then summarized by the following elegant form
for the $n$-point function on the disk
$$
D_{\alpha _{1} \ldots \alpha _{n}}={\rm Tr}_{\rm
\scriptstyle reg}\left( \varpi_{\alpha _{1}}\times
\cdots \times
\varpi_{\alpha _{n}}\right)
                                    \; .   \eqno (12)
$$
This rewriting suggests that the algebra ${\cal A}$
encodes all information about the underlying topological
vertex model. Indeed, as we will explicitly confirm below,
all correlation functions can be expressed in terms of the
structure constants
$P_{\alpha\beta}^{ \; \; \; \; \gamma}$ alone.
The alert reader will have, in fact, recognized that
${\cal A}$ plays the same role for {\it open} strings, as
the  algebra
${\cal H}$ did for closed ones. Indeed, the
link-flip condition is the condition of {\it planar}
duality, while the lack of commutativity reflects the
fact that open-string states cannot be freely permuted on
the boundary. Classifying  topological vertex models turns
out, therefore, to be equivalent to classifying
Chan-Paton factors for open strings.
This problem  was analyzed some time ago by Marcus and
Sagnoti [16]\footnote{$^{(*)}$}{\sevenrm These same
authors also suggested [17]  representing certain
Chan-Paton factors with boundary fermions
$\scriptstyle \psi^I$. These
can be considered as the remnant of a topological action,
$\scriptstyle \int d^{2}\!\sigma\,
\epsilon^{ij}\, \partial_i\psi^I\partial_j\psi^I$, on the
world-sheet.} and
we will essentially repeat their argument
below. Of course any open-string algebra ${\cal A}$ has a
closed-string descendant ${\cal H}$, defined by the two-
and three-point functions of the corresponding topological
vertex model on the sphere. The precise relation between
${\cal A}$ and  ${\cal H}$ will be established in
section~4.
\vskip 0.6cm
\noindent {\bf 3. Examples}
\vskip 0.3cm
Let us, however, first illustrate the above discussion with three concrete
examples of topological vertex models. These will, in particular, give us a
better understanding of how the advocated cloning of string theories
occurs.
\vskip 0.3cm
{\sl (a) Ferromagnetic Potts model.}
Consider Potts spins $i,j,\ldots \in \{1,\ldots , d\}$
located at the sites of
the triangular lattice and interacting among nearest
neighbours, as illustrated in
fig. 4. This model can be described in our language
by assigning to each oriented link a pair of indices that
designate the values of the spins at its endpoints,
$\alpha \equiv(i,j)$,
and by
choosing the link and plaquette weights as follows:
$$
\eqalign{ \ell^{(ij)(k\ell )} = &\,
\delta ^{jk\phantom )}
\delta^{\ell i\phantom )} \cr
P_{(ij)(k\ell )(mq)}   = \,
\delta_{jk\phantom )}
\delta_{\ell m\phantom )}
\delta_{qi\phantom )}&
W(i,j)\, W(k,\ell )\, W(m,q) \; ,\cr}
                                            \eqno(13)
$$
so as to ensure that all spins at a common endpoint
coincide.
Here
$W^2\!(i,j)=\kappa \exp{\delta _{ij}-1\over kT}$
is the Boltzmann weight for the corresponding link, with
$\kappa $ a constant.
Note that, because of the fluctuating coordination number
of vertices,  we cannot accomodate into $W$ a {\it constant
external field}, a remark that will play a role in the
sequel.
\vskip 0.3cm
Simple inspection shows that the Potts model satisfies the
topological invaria- nce conditions (10a, b) only in the
two extreme cases:  $W(i,j)=\delta_{ij}$ or
$W(i,j)=d^{-1/6}$,
corresponding to a $T=0 $ or $T\to \infty$ ferromagnet.
At $T=0$ all correlation functions on a connected surface
vanish unless all spins on all boundaries are aligned.
Thus there is one propagating open- and one closed-string
state for every value of the spin, or equivalently for
every world-sheet ground state. The corresponding
algebras are isomorphic to the trivial algebra of $d$
mutually annihilating  orthonormalized
idempotents:
${\cal A}\simeq{\cal H}\simeq {\bf C}^d$.
All diagonal couplings $\lambda_i = 1$, so
this model cannot distinguish the genus of the surface.
\vskip 0.3cm
At $T\to \infty$, on the other hand,
${\cal A}\simeq {\rm End}\left(
{\bf C}^{d}\right)$
is isomorphic to the
algebra of all $d\times d$ complex matrices. Since the
high-temperature phase is, however, unique, there is no
cloning of theories, {\it i.e.} there is a single propagating
closed-string state. Its coupling constant is
$\lambda=1/d$,
as can be read off from the partition function
$Z^{(\Gamma )}(T\to \infty)= d^{ n_2 - n_1 + n_0} =
d^{ 2-2\Gamma }$
where $n_2$, $n_1$ and $n_0$ are the
numbers of faces, edges and vertices of
${\cal M}_{\Gamma}$.
This is of course the well-known counting
used in the topological expansion of matrix models [14].
\vskip 0.3cm
{\sl (b) Lattice gauge theory.}
The link variables $f,g,h, \ldots$ are in this case
elements of a compact group $G$. The theory is
defined by
$$
\ell^{fg}=\delta (fg)
                                  \eqno(14{\rm a})
$$
and
$$
P_{fgh} =\sum_{r}  d_r\,
p_r \,  \chi_r(fgh)
                               \; ,\eqno(14{\rm b})
$$
where
$\delta (g)$
is the group
$\delta $-distribution, which
sets its argument equal to the identity.
The choice of link weights ensures that inverting
orientation amounts to group inversion. The
cyclically-symmetric plaquette weight, on the other hand,
is an arbitrary class function of the corresponding Wilson
loop, expressed in terms of its character decomposition.
Here  $d_r$ is the dimension of the  representation $r$
and $p_r$ are arbitrary coefficients. Using
the orthonormality relations
$$
\int_{G}dg\, R^{(r)}_{ij}\big(g\big)\,
R^{(r')}_{k\ell}\big(g^{-1}\big)=
{1\over d_{r}} \, \delta ^{rr'\phantom {)}}_{\phantom l}
\! \delta ^{\phantom {(r)}}_{jk}
\delta ^{\phantom {(r)}}_{\ell i}
                            \; ,           \eqno (15)
$$
where $dg$ stands for the normalized Haar measure and
$R^{(r)}_{ij}(g)$
for the matrix element of
$g$
in the  representation $r$, one can check that
$$
P_{fg}^{ \; \; \; h}\,
P_{hf'g'}^{\phantom g}  =
P_{gf'}^{ \; \; \; \, h} \,
P_{fhg'}^{\phantom g}=
\sum_{r}  d_{r}^{\phantom 2}\,
p_r^{2} \,  \chi _{r}^{\phantom 2}\big(fgf'g'\big)
                                   \; ,     \eqno (16)
$$
so that the link-flip move condition is automatically
satisfied. As a result, Yang-Mills in two
dimensions is, modulo area dependence, a topological
theory [12, 18].
Eq. (15) is actually all one needs to calculate
an arbitrary correlation function, with the result
[12]
$$
C^{\, (\Gamma )}_{\{g_{1}\}\cdots \{g_{n}\}}=
\sum_{r}d_{r {\phantom \}}}^{2-2\Gamma -n}\,
p_{r {\phantom \}}}^{A \phantom )}
\chi _{r {\phantom \}}}^{\phantom )}\! \!
\big(g_{1}^{\phantom \}}\big) \ldots
\chi _{r {\phantom \}}}^{\phantom )}\! \!
\big(g_{n}^{\phantom \}}\big)
                             \; ,          \eqno (17)
$$
where $g_{h}$ is the Wilson-loop
around the $h\, $th hole.  To prove this statement, one
deforms a surface by gluing  triangular  plaquettes on
its boundary and integrating out the group variable(s)
along the common links.
Each plaquette will contribute an extra power of $p_r$ in
the character decomposition of the result. The powers of
$d_r$ in eqs (14b) and (15) on the other hand, will
generically cancel out except when
one creates or destroys a connected component
of the boundary, in which case one loses or
gains, respectively, an extra power of $d_r$.
Since any Riemann surface can be
constructed this way, the above result
follows easily. In the case of a finite group of order
$\vert G \vert$
the above analysis holds, provided we take
$\ell ^{fg}={1\over  \vert G \vert}\, \delta (fg)
\equiv {1\over  \vert G \vert}\, \delta ^{fg,1}$
instead of (14a), and replace in
the orthonormality relations (15) the Haar measure by the
average over the group,
${1\over \vert G \vert}\sum_{g\in G}$.
\vskip 0.3cm
For the theory to be truly topological, the
correlation-functions must be area-independent. This
implies that for all  representations, either
$p_r = 0 $ or $p_r=1$\footnote{$^{(*)}$}{\sevenrm It is
amusing to observe that by choosing, for a
continuous group, the
coefficients  $\scriptstyle p_r$
appropriately, one can modify the string-susceptibility
exponent. We thank E.~Kiritsis for bringing up  this
point.}. Let us assume, in particular, that
$p_{r}=1 $ $\forall r$,
so that $P_{fg}^{ \; \; \; h}= \delta\big(fgh^{-1}\big)$.
In this case
${\cal A}$
is the so called  {\it group algebra}, and
the partition function measures
the volume of {\it flat} gauge connections on
${\cal M}_{\Gamma }$,
with the result:
$Z^{(\Gamma )}_{\phantom r} = \sum_r
d_r^{2-2\Gamma \phantom )}$. Furthermore, as seen from eq.
(17), inequivalent closed-string states are in one-to-one
correspondence with the {\it conjugacy classes} of the
group. Their algebra
${\cal H}$, read off from their two- and
three-point functions on the sphere, is isomorphic to the
so called {\it algebra of classes}. Comparing eqs (6) and
(17), we see that this algebra can be diagonalized in the
basis of representations
$\left\{ \tilde\phi_r\right\}$,  with diagonal structure
constants
$\lambda_r = 1/d_r$
\footnote{$^{(\dag)}$}{\sevenrm To avoid confusion, we
stress again that the diagonal structure constants are
meaningful because we normalized the propagator or
two-point function on the sphere to one.}.
We thus obtain one decoupled theory
for every irreducible representation of the group.
The change of basis between conjugacy classes and
representations is effected, for a finite
group, by the matrix
$S_{\; \, a}^{r}=\left\vert
{\cal C}_{a}^{\phantom r} \right\vert
\chi _{r}^{\phantom a}(g_{a}^{\phantom r})$,
where $\left\vert {\cal C}_{a} \right\vert$ is the number
of elements of the class
${\cal C}_{a}$ with representative $g_{a}$.
\vskip 0.3cm
{\sl (c) Gauged $\sigma $-model.}
A question that arises is whether we can find a model for
which ${\cal H}$ is the fusion algebra of group
representations, equipped with the usual conjugation
operation. Let us assume for definiteness that the group
$G$ is finite. Writing the fusion coefficients in the form
$$
{\cal N}_{r_1 \ldots r_n} =
\sum_{a}
{\left\vert {\cal C}_{a} \right\vert\over \vert G\vert}
\, \chi_{r_1}(g_a) \ldots \chi_{r_n}(g_a)
                                    \; ,          \eqno(18)
$$
shows that they are diagonalized in a basis of
classes, with diagonal couplings given by
$\lambda _{a}=\sqrt{\vert G\vert
/\left\vert {\cal C}_{a} \right\vert}$.
Let us therefore
try to construct a spin model like the one shown in
fig.~4, but with spins taking their values in $G$, and
with Boltzmann weights forcing them to lie in the same
conjugacy class
$$
W(f,h)=
\sum_{g\in G}\delta \big( fgh^{-1}g^{-1}\big)
=\sum_{r}\chi _{r}\big(f\big)\,
\chi _{r}\big(h^{-1}\big)
                    \; .          \eqno (19)
$$
This model is invariant
under spin transformations $f\to gfg^{-1}$, with $g$
chosen independently at every site. It resembles, in this
sense, the topological gauged WZW models $G/G$ [10], for
which ${\cal H}$ is believed to be the fusion algebra  of
the corresponding quantum group representations. This is,
however, where the analogy ends. Indeed, although the
closed-string algebra of the above spin model
{\it is} diagonalized in a basis of classes, its
diagonal couplings, after
appropriate normalizations, turn out to be
$\lambda _{a}=1/\left\vert {\cal C}_{a} \right\vert$,
in disagreement with (18). As we will in fact now
prove, a generic fusion algebra cannot be obtained from a
topological vertex model, because the inverse couplings of
the latter are necessarily quantized.
\vskip 0.6cm
\noindent {\bf 4. Classification}
\vskip 0.3cm
The classification of topological vertex models is a
simple corrolary of some standard results
on the structure of algebras [19], which we will use
without proof in this section. First let us suppose that
the algebra
${\cal A}$
associated with the topological
model has a non-trivial ($\not= \{0\}$) radical
${\cal R}$,
{\it i.e.} a non-trivial
maximal nilpotent two-sided
ideal. Being nilpotent, the elements of ${\cal R}$ have a
vanishing trace in any representation of the algebra.
Furthermore, being an ideal, ${\cal R}$ is closed under
multiplication by an arbitrary element of ${\cal A}$. Thus
for any
$\varrho \in {\cal R}$, we have
${\rm Tr}_{\rm \scriptstyle
reg}\left(\varpi_{\alpha } \times \varpi_{\beta } \times
\varrho \right)=0$  $\forall \alpha ,\beta$,
so that elements of the radical cannot
appear on any elementary plaquette of the surface. We may
therefore define an equivalent topological model by
restriction: ${\cal A} \to {\cal A}/{\cal R}$.  Put
differently we can assume without loss of generality that
${\cal A}$ has no radical and is hence {\it semi-simple}.
\vskip 0.3cm
Now a semi-simple algebra has a unique decomposition into
a direct sum of mutually annihilating simple components:
$$
{\cal A}=\bigoplus_{a=1}^{M}{\cal A}_{a}\; \;
{\rm with}\; \;
{\cal A}_{a}\times {\cal A}_{b}=0\; \;
{\rm if }\; \;
a\not=b\; .
                                       \eqno (20)
$$
A simple algebra over the complex field,
on the other hand,
is always
isomorphic to a complete matrix algebra
$$
{\cal A}_{a}\simeq {\rm End}\left(
{\bf C}^{d_{a}}\right)\; .
                                       \eqno (21)
$$
Thus the most general ${\cal A}$ is isomorphic to an
algebra of all block-diagonal matrices, with $M$ blocks of
sizes $d_a\times d_a$ ($a=1,\ldots, M$); the dimension of
${\cal A}$ is
$s=\sum_{a=1}^{M}d_{a}^{2}$. For instance
the group algebra is isomorphic to
$\bigoplus_{r}  {\rm End}\left( {\bf C}^{d_{r}}\right)$,
while its dimension, for a finite group, is precisely the
number of group elements,
$\sum_{r}d_{r}^{2}=\vert G\vert$.
\vskip 0.3cm
The $M$  components in the decomposition (20)
correspond to $M$ completely decoupled theories.
Indeed, from (12) we see that the plaquette weights,
and hence also
all other correlation functions, vanish
unless all boundary spins belong to the same irreducible
component. Let us therefore concentrate
on a single component or, to simplify notation, take
${\cal A}={\rm End}\left({\bf C}^{d}\right)$.
We may choose for this algebra the basis
$\{\tau _{\alpha }\; ,\; \; \alpha =1,\ldots, d^{2}\}$
of $d\times d$ matrices, such that
$\tau _{1}={1\over d}\, {\bf 1}_{d\times d}$,
while the remaining $\tau _{\alpha}$
are hermitean, traceless  and normalized so that
${\rm Tr}(\tau _{\alpha }\tau _{\beta })={1\over d}\,
\delta _{\alpha \beta }$.
These matrices provide in fact
the only non-trivial irreducible representation of
the algebra, contained $d$ times inside
the regular representation. From the expression (12) for
the two- and three-point functions on the disk it is then
straightforward to deduce that
$$
\eqalign{
&\ell ^{\alpha \beta }= \delta ^{\alpha \beta } \cr
P_{\alpha \beta \gamma }&=d\,
{\rm Tr}(\tau _{\alpha } \tau _{\beta } \tau _{\gamma })
                                    \; .
\cr}                                         \eqno (22)
$$
Note in particular that
$\ell _{\alpha \beta }$
is nothing but the two-point function on the disk,
$D_{\alpha \beta }$, which is invertible because
${\cal A}$
has no radical.
Substituting now into expression (8) for the two-point function on the sphere,
and
using the orthocompleteness relation
$$
\sum_{\alpha =1}^{d^{2}}
(\tau _{\alpha })_{ij}\,
(\tau _{\alpha })_{k\ell }=
{1\over d}\, \delta _{jk}\,
\delta _{\ell i}   \; \;
\forall i,j,k,\ell \in \{1, \ldots, d\}
                            \; ,  \eqno (23)
$$
one finds
$$
C_{\alpha \beta }={\rm Tr}(\tau _{\alpha })\,
{\rm Tr}(\tau _{\beta })
                                   \; .    \eqno (24)
$$
Since ${\rm Tr}(\tau _{\alpha })= \delta_{\alpha 1}$, we
conclude that, out of the
$d^2$
open-string states in ${\cal A}$, {\it only the one
corresponding to the identity propagates as a
closed-string state}. Allowing holes of length more than
one does  not, in fact, introduce any new
linearly-independent states in ${\cal H}$. Indeed, with
the help of eqs (22) and (23) we can compute an arbitrary
correlation function with the result
$$
C_{\left\{\alpha ^{(1)}\right\}\cdots \left\{\alpha
^{(n)}\right\}}^{\, (\Gamma )}
=
d_{\phantom {\left\{\alpha ^{(1)}\right\}}}
^{2-2\Gamma-n \phantom {)}}
\prod_{h=1}^{n}{\rm Tr}\left(
\tau ^{\phantom a}_{\alpha ^{(h)}_{1}
}\ldots
\tau ^{\phantom a}_{\alpha ^{(h)}_{\ell_{h}}
}
\right)
                            \; ,    \eqno (25)
$$
where
$\ell_{h}$ is the length of the $h\,$th hole, and
$\left\{\alpha ^{(h)}_{\phantom 1}\right\} =
\left\{\alpha ^{(h)}_{1},\ldots ,
\alpha ^{(h)}_{\ell_{h}}\right\}$ the
values of the spins ordered around it modulo cyclic
permutations. Clearly any closed string state
$\left\{\alpha ^{(h)}\right\}$
is a simple multiple of the state
$\{1\}$,  the three-point coupling of this latter being
equal to ${1/d}$. We thus conclude that the only
effect of adding a topological vertex model with
${\cal A}= {\rm End}\left( {\bf C}^{d}\right)$ to the
matter content of a closed string theory is to renormalize
the string coupling constant. More generally, the space
of propagating closed-string states is the center of
${\cal A}$, and contains $M$ elements corresponding to the
identities of the simple components.
\vskip 0.3cm
Both this renormalization by an inverse integer, and the
cloning of theories when ${\cal A}$ has more than one
simple component, have been already illustrated by the
infinite- and zero-temperature Potts ferromagnet. The
point of our discussion here was to prove that an {\it
arbitrary} topological vertex model can be reduced to the
above two simple effects. We summarize this in the
following
\vskip 0.3cm
{\sl PROPOSITION.$ \, -\, $}{\it Inequivalent
topological vertex models are in one-to-one correspondence
with semi-simple algebras ${\cal A}$ over the complex
numbers. These, in turn, are isomorphic to a direct sum of
$M$ complete matrix algebras of dimensions $d_{a}^{2}$ for
$a=1,\ldots , M$.  The center of ${\cal A}$ is the
commutative algebra of closed-string states, ${\cal H}$.
The diagonalized
structure constants of this latter in an orthonormal
basis are
$\lambda _{a}^{\phantom 1} = 1/d_{a}$.}
\vskip 0.3cm
The above analysis goes in fact through with little
change, if one drops the constraint
of invariance under the pyramid move of fig. 3({\it b}).
The net effect is that for every simple component of
${\cal A}$, $\ell _{\alpha \beta }$ can now be an
arbitrary multiple of the identity, {\it i.e.} the
two-point function
$D _{\alpha \beta }$
on the disk of area two, eq. (7).
This corresponds to an
independent renormalization of the world-sheet
cosmological constant for every decoupled component of the
theory as illustrated in the example of gauge theory
with arbitrary plaquette action, eq. (17).
\vskip 0.3cm
More important is the fact that the above
classification was based on the assumption of
equivalence under the complex transformations (9). These
allow us, for example, to transform lattice gauge theory
into an appropriate Potts model. Nevertheless, it could
happen that only a subset of these transformations are
admissible. Consider for instance the coupling of
topological matter to pure gravity, described by a
matrix model with partition function
$$
{\cal Z}_{\rm grav\ +\ top\ mat}=
\int \prod_{\alpha =1}^{s}\!d\Phi^{\alpha }\,
e^{-{\rm Tr}\left(
\ell _{\alpha \beta }\, \Phi ^{\alpha }\Phi ^{\beta }-
{\mu \over \sqrt{N}}P_{\alpha \beta \gamma }\,
\Phi ^{\alpha }\Phi ^{\beta }\Phi ^{\gamma }
\right)}
  \; ,                                     \eqno (26)
$$
where $\Phi^{\alpha }$
are hermitean $N\times N$ matrices,
$1/N$ is the bare string coupling constant and $\ln \mu $
the world-sheet bare cosmological constant.
The perturbative expansion of this integral is left
invariant under arbitrary complex changes of basis,
$\Phi^{\alpha }_{\phantom \beta }\to  T^{\alpha
}_{\; \, \beta }\,  \Phi^{\beta}_{\phantom \alpha }$.
As a result, we conclude from our discussion that
within perturbation theory  the model (26) is equivalent
to  $M$ decoupled models of hermitean matrices with
renormalized sizes  $(d_a N)\times (d_a N)$.  Note,
however, that only  {\it real} transformations leave
invariant the integration contours in the
complex-$\Phi^{\alpha}$ space. This could be important if
one had a non-perturbative definition of the integral.
Another situation in which only real transformations are
allowed, is when one wants to twist by a parity  that
distinguishes hermitean and anti-hermitean states of the
topological model. In such circumstances we need the
finer classification
of algebras over the field ${\bf R}$ of real numbers. A
generic ${\cal A}$ is now a direct sum of complete matrix
algebras over a finite extension of ${\bf R}$,  {\it i.e.}
over the  real,
complex or quaternion fields. This is precisely what
allows the addition of $SO(d)$, $U(d)$ or $U\! Sp(d)$
quantum numbers to open strings with Chan-Paton factors
[16].
\vskip 0.6cm
\eject
\noindent{\bf 5. Conclusions}
\vskip 0.3cm
We conclude our brief tour of topological vertex
models with two remarks. The first concerns the
quantization of the inverse couplings of closed-string
states in the diagonal ("tilde") basis. This quantization
is not required by the axioms of ref. [3], and there are
examples, such as the coset $G/G$ models [10], for which
it does not hold. Such models could, to be sure, be put on
the lattice if we allowed a constant external field to act
on the spins in fig. 4, or else [20] if spins could take
their values in a quantum group  {\it {\`a} la} Woronowicz
[21]. Neither construction seems, however, compatible
with a local, cyclically symmetric plaquette action, or
equivalently with a factorizable open-string ascendant.
The question of how to couple such models to a theory of
both open and closed strings deserves further study.
\vskip 0.3cm
Our second comment concerns the wormhole of
fig. 1. We may think of this as describing the approach to
the simplest topological model, namely the Ising model at
zero temperature. Indeed, consider a background for which
the temperature
$T$ varies continuously from large values to zero as a
function of Euclidean time. Then a splitting of worlds
will occur precisely when $T$ crosses the critical
temperature into the ferromagnetic phase. Alternatively,
we may consider a cosmological scenario in which the
conformal factor, or the distance on the triangulated
world-sheet, plays the role of Minkowski time [22]. If the
Ising temperature were chosen below criticality at the
cut-off, it would renormalize towards zero at larger
scales, so that two decoupled worlds would emerge
asymptotically in time. It may be possible to study this
quantitatively, by considering the  appropriate flows in
the two-matrix model.
\vskip 0.6cm
\centerline{\bf Acknowledgements}
\vskip 0.3cm
We thank D.~Boulatov, E.~Kiritsis, J.~Lascoux,
J.-M.~Maillet, V.~Pasquier, P.~Roche
and A.~Sagnotti for discussions as well as C.~Itzykson
for a careful reading of the manuscript. One of
us (C.B.) aknowledges travel support from EEC grant
SC1-0394-C.
\vskip 0.6cm
\centerline{\bf References}
\vskip 0.3cm
\item{[1]}{E. Witten, Commun. Math. Phys. {\bf 117} (1988)
353 and {\bf 118} (1988) 411.}
\vskip 0.05cm
\item{[2]}{For a recent review see D. Birmingham, M.
Blau, M. Rakowsky and G.~Thompson, Phys. Rep. {\bf 209}
(1991) 129; also R. Dijkgraaf, E. Verlinde and
H.~Verlinde in {\sl Random surfaces, quantum gravity and
strings}, proceedings of the 1990 Carg{\`e}se Workshop,
O.~Alvarez {\it et al.}~(eds), Plenum Press, New York
1991.}
\vskip 0.05cm
\item{[3]}{M. Atiyah, Publ. Math. I.H.E.S. {\bf 68}
(1988) 175; R. Dijkgraaf, Ph.D. Thesis, University of
Utrecht 1989.}
\vskip 0.05cm
\eject
\item{[4]}{D. Montano and J. Sonnenschein, Nucl. Phys.
{\bf B313} (1989) 258; R. Dijkgraaf and E. Witten,
Nucl. Phys. {\bf B342} (1990) 486; K. Li, Nucl. Phys.
{\bf B354} (1991) 711 and 725; R. Dijkgraaf, E. Verlinde
and H. Verlinde, Nucl. Phys. {\bf B352} (1991) 59.}
\vskip 0.05cm
\item{[5]}{C. Bachas and P.M.S. Petropoulos, Phys. Lett.
{\bf 247B} (1990) 363; E.~Witten,
IASSNS-HEP-90/45 preprint, (May 1990).}
\vskip 0.05cm
\item{[6]}{E. Witten, Phys. Rev. {\bf D44} (1991) 314.}
\vskip 0.05cm
\item{[7]} {E. Witten, Phys. Rev. Lett. {\bf 61} (1988)
670.}
\vskip 0.05cm
\item{[8]}{L. Baulieu and I.M. Singer, Commun.
Math. Phys. {\bf 125} (1989) 227.}
\vskip 0.05cm
\item{[9]}{T. Eguchi and S.-K. Yang,
Mod. Phys. Lett. {\bf A5} (1990) 1693.}
\vskip 0.05cm
\item{[10]}{K. Gawedzki and A. Kupianen, Nucl. Phys. {\bf
B320} (1989) 625; E.~Witten, IASSNS-HEP-91/26 preprint,
(June 1991); M. Spiegelglas and S. Yankielowicz,
Technion PH-34-90 preprint, (January 1992).}
\vskip 0.05cm
\item{[11]}{J. Wheater, Phys. Lett. {\bf
223B} (1989) 451; T. Jonsson, Phys. Lett. {\bf 265B}
(1991) 141; T.~Filk, UF-THEP 92/1 preprint, (January
1992).}
\vskip 0.05cm
\item{[12]}{E. Witten, Commun. Math. Phys.
{\bf 141} (1991) 153; R. Dijkgraaf and E.~Witten, Commun.
Math. Phys. {\bf 129} (1990) 393.}
\vskip 0.05cm
\item{[13]}{More recently this was discussed by
S. Elitzur, A. Forge
and E. Rabinovici, CERN-TH.6326 preprint, (November
1991).}
\vskip 0.05cm
\item{[14]}{J. Ambj{\o}rn, B. Durhuus and J.
Fr{\"o}hlich, Nucl. Phys.  {\bf B257 [FS14]} (1985) 433;
F. David, Nucl. Phys. {\bf B257 [FS14]} (1985) 45 and 543;
V.A. Kazakov, Phys. Lett. {\bf 150B} (1985) 282; V.A.
Kazakov, I.K. Kostov and A.A. Migdal,  Phys. Lett. {\bf
157B} (1985) 295.}
\vskip 0.05cm
\item{[15]} {J.W. Alexander, Ann. Math.
{\bf 31} (1930) 292; D.V. Boulatov, V.A. Kazakov, I.K.
Kostov and A.A. Migdal, Nucl. Phys. {\bf B275 [FS17]}
(1986) 641.}
\vskip 0.05cm
\item{[16]}{N. Marcus and A. Sagnotti, Phys.
Lett. {\bf 119B} (1982) 97; J.H.~Schwarz, proceedings
of the Johns Hopkins Workshop (1982) 233.}
\vskip 0.05cm
\item{[17]}{N. Marcus and A. Sagnotti, Phys.
Lett. {\bf 188B} (1987) 58.}
\vskip 0.05cm
\item{[18]}{S.G. Rajeev, Phys. Lett. {\bf 212B} (1988)
203;  D.S. Fine, Commun. Math.
Phys. {\bf140} (1991) 321.}
\vskip 0.05cm
\item{[19]}{B.L. van der Waerden {\sl Modern algebra}
{\bf 2}, Frederick Ungar Publ. Co., New~York 1949;
Th. Kahan {\sl Th{\'e}orie des groupes en physique
classique et quantique}, Dunod, Paris 1960.}
\vskip 0.05cm
\item{[20]}{D. Boulatov, private communication.}
\vskip 0.05cm
\item{[21]}{S.L. Woronowicz, Commun. Math. Phys.
{\bf 111 } (1987) 613.}
\vskip 0.05cm
\item{[22]}{J. Polchinski, Nucl. Phys. {\bf B324}
(1989) 123; S.R. Das,
S. Naik and S. Wadia, Mod. Phys. Lett. {\bf A4}
(1989) 1033; I. Antoniadis, C.P. Bachas,  J.~Ellis and
D.V.~Nanopoulos,  Phys. Lett. {\bf 211B} (1988) 393; Nucl.
Phys. {\bf B328} (1989) 117.
}
\vskip 0.6cm
\eject
\centerline{\bf Figure captions}
\vskip 0.3cm
\item{FIG. 1}{A
non-perturbative contribution to the functional
integral for the wavefunction of two universes.
Perturbative contributions would correspond to
(Euclidean) space-time histories with the topology of
two disjoint disks.}
\item{FIG. 2}{({\it a}) The elementary
area-$2$ disk defining the correlator
$D_{\alpha \beta }$.}
\item{}{({\it b}) The elementary
area-$2$ cylinder defining the correlator
$C_{\alpha \beta }$.}
\item{FIG. 3}{({\it a}) The link-flip move.}
\item{}{({\it b}) The pyramid move.}
\item{FIG. 4}{Conventional spin model on the
triangular lattice.
This model is topological only at  $T\to \infty $ or,
for ferromagnetic interactions, $T=0$.}
\vfil

\bye